\documentclass[aps,pra,floatfix,twocolumn,showpacs]{revtex4}
\usepackage{epsfig}
\usepackage{graphicx}
\usepackage{dcolumn}
\begin{document}
\title{Inconsistencies between lifetime and polarizability measurements in Cs}
\author{M. S. Safronova}
 \email{msafrono@physics.udel.edu}
\altaffiliation[Current address: ]{Department of Physics and Astronomy, 
University of Delaware, Newark, Delaware, 19716}
\author{Charles W. Clark}
 \affiliation{
Electron and Optical Physics Division,
National Institute of Standards and Technology,
Technology Administration,
U.S. Department of Commerce, 
Gaithersburg, Maryland 20899-8410}
\date{\today} 
\begin{abstract}
Electric-dipole matrix elements for $6p-nd$, $n=5$, $6$, $7$ 
transitions in cesium are calculated using
a relativistic all-order
method. The resulting matrix elements are used to evaluate $5d$ lifetimes and 
$6p$ polarizabilities. The data are compared with experimental
 lifetime and polarizability measurements made by different groups. 
Domination of the $6p$
scalar polarizabilities  
by $5d-6p$ dipole matrix elements 
facilitates an exacting consistency check of  
$5d$ lifetime and $6p$ polarizability data. 
Values of $5d-6p$ matrix elements
obtained from experimental $5d$ lifetime data are 
found to be inconsistent with those inferred from 
$6p$ polarizabilities derived from experimental Stark
shift data.  Our {\it ab initio} calculated $6p$
polarizabilities agree well with experimental determinations.    
\end{abstract}
\maketitle

The understanding of the accuracy of {\it ab initio} calculations
 in cesium is vital for the analysis of the Cs parity nonconservation (PNC) experiment \cite{CsPNC}.
In 1999, motivated by a number of recent
high-precision experiments, Bennett and Wieman \cite{beta} reanalyzed the 
agreement of theoretical calculations and experimental data for a 
number of Cs atomic properties and reduced the
previous theoretical uncertainty in the PNC amplitude by a factor of two.
Utilizing  measurements of the tensor transition polarizability,
$\beta$, reported in same work, they demonstrated a 2.5$\sigma$ 
discrepancy between the value of the 
weak charge $Q_W$ predicted by the Standard Model and that derived from the Cs PNC experiment.
Although several papers (for example, \cite{ref5,ref4,ref3,al,new1,new3,cwc}),  
have addressed this disagreement since 1999, 
the issue of the accuracy of {\it ab initio} calculations in Cs continues to
be of interest. 

In this work, we investigate the radiative properties of Cs $6p - nd$ transitions.  Although these do not bear directly on PNC experiments done to date, they have been the subject of careful experimental investigation, and thus provide benchmarks for precise comparison of theory and experiment.  In particular, there exist two independent measurements of the lifetimes of the $5d$ states \cite{5dT,5dH}, which do not agree within their stated uncertainties.  There also exist several experimental determinations of the $6p-6s$ Stark shifts which allow to infer the values of polarizabilities of the $6p$ states\cite{6pH,6pT,6pHH}.  Here we show that {\it ab initio} theory can check the mutual consistency of $5d$ lifetime and $6p$ polarizability data, with an accuracy of about 1\%.  We find the lifetime and polarizability
results to be inconsistent at this level.  Our calculations agree with the experimental values of $6p$ polarizabilities, but deviate from both determinations of the $5d$ lifetimes.  We suggest that further experiments
are desirable in order to clarify this issue.  In addition, understanding of the 
accuracy of the $5d$ state properties in Cs is germane to the ongoing 
PNC experiment in isoelectronic Ba$^+$ \cite{ba}, since the $5d$ state is directly involved in this experiment.

In outline, our approach uses a relativistic all-order method to calculate
electric-dipole matrix elements for Cs $6p-nd$ transitions for $n=5$, $6$, 
and $7$.  These are used to evaluate $5d$ radiative lifetimes and 
$6p$ polarizabilities (for the latter, we also include contributions
from all other relevant states).  
Our calculations of the $6p$ scalar polarizabilities,
which are in good agreement with experiment, show that 
they are dominated by
contributions from $5d-6p$ transitions. 
These are the only electric-dipole transitions contributing to the 
$5d$ state lifetimes (as we mention below, the $5d-6s$ electric
quadrupole transition rates are negligibly small). Thus, it 
is possible to check consistency
between polarizability and lifetime measurements by deriving $5d-6p$ matrix elements 
from $5d$ lifetime measurements and substituting these values into the $6p$ polarizability
calculations.  For either of the two experimental lifetimes,
\cite{5dT,5dH} this procedure yields a result that disagrees
with directly measured polarizabilities
\cite{6pH,6pT,6pHH} by several standard deviations.
 \begin{table*} [ht]
\caption{\label{tab1}Absolute values of electric-dipole $5d-6p$ reduced matrix elements 
in Cs calculated in different 
approximations:  Dirac-Hartree-Fock (DHF), third-order many-body perturbation theory (III),
 single-double all-order method (SD), single-double 
all-order method including  partial triple contributions (SDpT) 
and the corresponding scaled values.  $R$ is the ratio of the $5d_{3/2}-6p_{3/2}$ to $5d_{3/2}-6p_{1/2}$
transition matrix elements. All values are given in atomic units ($ea_0$, where $a_0$ is the Bohr radius).} 
 \begin{ruledtabular}  
\begin{tabular}{lrrrrrr} 
\multicolumn{1}{c}{Transition} &
\multicolumn{1}{c}{DHF} & 
\multicolumn{1}{c}{III} &
\multicolumn{1}{c}{SD} & 
\multicolumn{1}{c}{SD$_{\text{sc}}$} &
\multicolumn{1}{c}{SDpT} & 
\multicolumn{1}{c}{SDpT$_{\text{sc}}$} \\
\hline
$5d_{3/2}-6p_{1/2}$&  8.9784 &  6.9231 &   6.5809 &  7.0634&   6.9103 & 7.0127\\
$5d_{3/2}-6p_{3/2}$&  4.0625 &  3.1191 &   2.9575 &  3.1871&   3.1112 &	3.1614\\
$R$                &  0.4525 &  0.4505 &   0.4494 &  0.4512&   0.4502& 0.4508\\
$5d_{5/2}-6p_{3/2}$& 12.1865 &  9.4545 &   9.0238 &  9.6588&   9.4541 & 9.5906\\
 \end{tabular}   
\end{ruledtabular}
 \end{table*}        
 
The particular
all-order method used here is the linearized coupled-cluster method
which sums infinite sets of many-body perturbation theory terms.
We refer the reader to Refs.~\cite{sd,cs,relsd} for a detailed
description of the approach. 
The wave function of
the valence electron $v$ is represented as an expansion
\begin{eqnarray}
 |\Psi_v \rangle &= &\left[ 1 + \sum_{ma} \, \rho_{ma}
a^\dagger_m a_a + \frac{1}{2} \sum_{mnab} \rho_{mnab} a^\dagger_m
a^\dagger_n a_b a_a +
 \right. \nonumber \\
&+& \left. \sum_{m \neq v} \rho_{mv} a^\dagger_m a_v + \sum_{mna}
\rho_{mnva} a^\dagger_m a^\dagger_n a_a a_v \right. \nonumber \\
&+&  \left.\frac{1} {6}\sum_{mnrab} \rho_{mnrvab}
 a_m^{\dagger } a_n^{\dagger} a_r^{\dagger } a_b a_aa_v \right]|
\Phi_v\rangle  , \label{eq1}
\end{eqnarray}
where $\Phi_v$ is the lowest-order atomic state function, which is
taken to be the {\em frozen-core} Dirac-Hartree-Fock (DHF)
    wave function
 of a state $v$. This lowest-order atomic state function
can be written as $ |\Phi_v\rangle =a_v^{\dagger }|0_C\rangle,
$ where $|0_C\rangle $ represent DHF  wave function of a closed
core. In equation (\ref{eq1}), $a^\dagger_i$ and $a_i$ are
creation and annihilation operators, respectively. 
The indices $m$, $n$, and $r$ designate excited states and indices $a$ and $b$ designate 
core states.
The excitation coefficients $\rho_{ma}$, $\rho_{mv}$, $\rho_{mnab}$, and $\rho_{mnva}$
are used to calculate matrix elements, which can be expressed in the framework of the 
all-order method  as 
 linear or quadratic functions of the excitation coefficients. We restrict the expansion given by Eq.~(\ref{eq1}) to single and double (SD) excitations, with partial inclusion of triple excitations. The results obtained using the SD expansion 
are referred to as SD data throughout the paper and results obtained with partial addition of the triple excitations are referred to as SDpT data.  
We also performed third-order many-body perturbation
theory calculations, following Ref.~\cite{ADNDT},
to better understand the size of higher-order correlation corrections.
Unless stated otherwise, all results in this paper are expressed 
in the familiar system of atomic units, a.u., in
which unit values are assigned to the elementary charge, $e$, the mass of the
electron, $m$, and the reduced Planck constant $\hbar$. 
  
Table~\ref{tab1} lists the $5d-6p$ reduced electric-dipole matrix elements 
in Cs as calculated using the Dirac-Hartree-Fock approximation (DHF), third-order perturbation theory (III),
single-double all-order method (SD), and single-double 
all-order method including  partial triple contributions (SDpT).
We use semi-empirical scaling described, for example, in Ref.~\cite{cs} to estimate
some classes of the omitted high-order corrections. The scaled values are listed in rows
labeled SD$_{\text{sc}}$ and SDpT$_{\text{sc}}$. 
 \begin{table*}
\caption{\label{tab4} The values of 
Einstein A-coefficients $A_{vw}$ (in MHz) and final lifetimes (in ns) for $5d_{5/2}$
and $5d_{3/2}$  states in Cs. The theoretical values are compared with 
experimental results from \protect\cite{5dT} and \protect\cite{5dH}.}
 \begin{ruledtabular}
\begin{tabular}{lclcccccc}
\multicolumn{1}{c}{Level} &
\multicolumn{1}{c}{Transition} &
\multicolumn{1}{c}{} & 
\multicolumn{1}{c}{SD} & 
\multicolumn{1}{c}{SD$_{\text{sc}}$} &
\multicolumn{1}{c}{SDpT} & 
\multicolumn{1}{c}{SDpT$_{\text{sc}}$}&
 \multicolumn{1}{c}{Expt.~\protect\cite{5dT}} &
 \multicolumn{1}{c}{Expt.~\protect\cite{5dH}} \\
    \hline
 $5d_{5/2}$& $5d_{5/2}-6p_{3/2}$ & $A_{vw}$ &   0.646 & 0.741	&  0.710 &  0.730&&\\
	  &                     & $\tau$   &   1547  &	1350   &   1409 &   1369&1281(9)&1226(12)\\  \hline		  
 $5d_{3/2}$ &$5d_{3/2}-6p_{1/2}$&$A_{vw}$&   0.804 &   0.926  &   0.886 & 0.913 &&\\
          &$5d_{3/2}-6p_{3/2}$&$A_{vw}$&   0.094 &  0.109  &   0.104 & 0.107 &&\\
          &                   &$\tau$  &   1114  &  966    & 	1010  &    981 & 909(15)& \\	  
 \end{tabular}
\end{ruledtabular}
 \end{table*} 
   \begin{table}
\caption{\label{tab6a}  Contributions to the $6p_{1/2}$ and $6p_{3/2}$ scalar polarizabilities $\alpha_0$ in Cs and their uncertainties $\delta \alpha_0$, in
units of $a^3_0$. The values of corresponding matrix elements $d$ (in a.u.), their sources and uncertainties
  $\delta d$ (in \%) are also given. The $6p-6d$
 and $6p-7d$ matrix elements are from the present SDpT all-order calculation.}
 \begin{ruledtabular}
\begin{tabular}{lrrcrr}
 \multicolumn{1}{c}{$\alpha_0(6p_{1/2})$} &
\multicolumn{1}{c}{$d$} &
\multicolumn{1}{c}{$\delta d$} &
\multicolumn{1}{c}{} &
\multicolumn{1}{c}{$\alpha_0$} &
\multicolumn{1}{c}{$\delta \alpha_0$} \\
\hline
    $6p_{1/2}-5d_{3/2}$& -7.283   &    0.8&\protect~\cite{5dT}\footnotemark[1]& 1168.4&    18.7  \\
    $6p_{1/2}-6s      $& -4.489   &    0.1&\protect~\cite{6s6p}& -131.9&    -0.3  \\
    $6p_{1/2}-6d_{3/2}$&  4.145   &    4.8&SDpT&  110.2&    10.6  \\
    $6p_{1/2}-7s      $& -4.236   &    0.5&\protect~\cite{al}&   178.4&     1.8  \\
    $6p_{1/2}-7d_{3/2}$&  2.033   &    1.7&SDpT&   20.3&     0.7  \\
    $6p_{1/2}-8s      $& -1.026   &    0.6&\protect~\cite{al}&    5.9&     0.1  \\
    $6p_{1/2}-9s      $&  0.550   &    0.5&\protect~\cite{al}&    1.4&     0.0  \\
  $\alpha_{\text{tail}}$&         &       &DHF&   35.4&  10.6  \\
  $\alpha_{\text{core}}$&         &      &\protect~\cite{jkh}& 15.8& 0.3 \\
  Total &                           &    &&1404&24\\
 \hline
  \multicolumn{1}{c}{$\alpha_0(6p_{3/2})$} &
 \multicolumn{1}{c}{$d$} &
\multicolumn{1}{c}{$\delta d$} &
\multicolumn{1}{c}{} &
\multicolumn{1}{c}{$\alpha_0$} &
\multicolumn{1}{c}{$\delta \alpha$} \\ 
\hline
  $6p_{3/2}- 5d_{3/2}$&  3.286 &         0.9&\protect~\cite{5dT}\footnotemark[1] &  142.7 &   2.6  \\
  $6p_{3/2}- 5d_{5/2}$&  9.916 &         0.3&\protect~\cite{5dT} & 1255.5 &   8.8 \\
  $6p_{3/2}- 6s      $& -6.324 &         0.1&\protect~\cite{6s6p}& -124.7 &  -0.2  \\
  $6p_{3/2}- 6d_{3/2}$& -2.053 &         4.6&SDpT        &   14.2 &   1.3 \\
  $6p_{3/2}- 6d_{5/2}$& -6.010 &         4.3&SDpT       &  121.2 &  10.4 \\
  $6p_{3/2}- 7s      $& -6.473 &         0.5&\protect~\cite{al}  &  225.3 &   2.3 \\
  $6p_{3/2}- 7d_{3/2}$& -0.969 &         1.5&SDpT       &    2.4 &   0.1 \\
  $6p_{3/2}- 7d_{5/2}$& -2.868 &         1.4&SDpT        &   21.0 &   0.6 \\
  $6p_{3/2}- 8s      $& -1.462 &         0.6&\protect~\cite{al}    &    6.2 &   0.1  \\
  $6p_{3/2}- 9s      $&  0.774 &         0.6&\protect~\cite{al}    & 	 1.4 &   0.0 \\
$\alpha_{\text{tail}}$ &       &            &DHF&    38.7&   11.6\\
$\alpha_{\text{core}}$ &         &          &\protect\cite{jkh}&    15.8&  0.3\\
Total                 &          &          &&    1720 & 18	 
\end{tabular}
\end{ruledtabular}
\noindent \footnotetext[1]{Derived from the experimental $5d_{3/2}$ lifetime 
\protect~\cite{5dT} using  theoretical ratio
of the $6p_{3/2}- 5d_{3/2}$ and  $6p_{1/2}- 5d_{3/2}$ matrix elements. } 
 \end{table} 
 
 We use the $5d-6p$ matrix elements from Table~\ref{tab1} to calculate the lifetimes of the 
 $5d_{3/2}$ and $5d_{5/2}$ levels in Cs.  
  The Einstein A-coefficients $A_{vw}$ 
  are calculated using 
 the formula \cite{ADNDT}
 \begin{equation}
 A_{vw}=\frac{2.02613\times 10^{15}}{\lambda^3}
 \frac{|\langle v\|D\|w\rangle|^2}
 {2j_v+1}\,s^{-1},
 \label{eq33}
 \end{equation}
 where $\langle v\|D\|w\rangle$ is the reduced electric-dipole 
 matrix element for the transition between states $v$ and $w$ and   
  $\lambda$ is corresponding wavelength in nm. The lifetime
  of the state $v$ is calculated as
   \begin{equation}
\tau_v=\frac{1}{\sum_{w}A_{vw}}. \label{eq3} 
 \end{equation}
  The results are listed in Table~\ref{tab4}. The experimental 
 energies from \cite{exp} are used. 
 The scaled SD values are taken as final values based on the comparison of a number of Rb, Cs, and Fr results 
 \cite{th,rb} with experiment.  
  The theoretical values differ substantially, by over 5\%,
from the experimental results (we note that that the experimental values from Refs.~\cite{5dT,5dH}
differ by 4\%, which exceeds their stated 
uncertainties of 0.7\% and 1\%, respectively).  One possible source of such a discrepancy 
is the contribution of the $5d-6s$ electric-quadrupole transition to the $5d$ lifetime. Our calculation of this rate, using the all-order 
method, yields a corresponding Einstein A-coefficient for the $5d_{5/2}-6s$ transition of 19~Hz, which 
is only 0.02\% of the corresponding electric-dipole 
A-coefficient of 741~kHz (see Table~\ref{tab4}). Thus, the contribution of the electric-quadrupole transition to $5d$ lifetime is entirely negligible within
the present experimental and theoretical uncertainties.
 
 To clarify such a large disagreement we check the consistency of the
experimental $5d$ lifetime measurements with 
$6p$ polarizability measurements, which involves contributions from the same 
transitions. First, we use experimental $5d$ lifetimes from \cite{5dT} to determine
the $5d-6p$ reduced matrix elements.
Inverting Eq.~(\ref{eq3}), we find for the  $5d_{5/2}-6p_{3/2}$ matrix element:
\begin{equation}
 |\langle 5d_{5/2}\|D\|6p_{3/2}\rangle|=9.916(35).
 \end{equation}
  To derive the $5d_{3/2}-6p_{1/2}$ and  $5d_{3/2}-6p_{3/2}$ matrix element, the lifetime of the $5d_{3/2}$
  level alone is not sufficient and some assumption about the ratio $R$
  of these matrix elements must be made.
 We use the theoretical SD$_{\text{sc}}$ value 0.4512(18) from Table~\ref{tab1}
  for the ratio and assume the 
 deviation of other high-precision theoretical results in Table~\ref{tab1} from this value to be its uncertainty. 
 The variation
 of the ratio from one approximation to another is far smaller than the
  variation in the individual 
 matrix elements, thus the uncertainty is rather low (0.4\%). 
 The resulting values of the $5d_{3/2}-6p$ matrix elements are:
 \begin{eqnarray}
     |\langle 5d_{3/2}\|D\|6p_{1/2}\rangle|&=&7.283(60),\nonumber \\
      |\langle 5d_{3/2}\|D\|6p_{3/2}\rangle|&=&3.286(27)(13). 
   \end{eqnarray}
   We separated the uncertainties in the $5d_{3/2}-6p_{3/2}$ matrix elements 
into contributions from
the $5d_{3/2}$ lifetime measurement (0.027) and from the estimation of $R$ (0.013).  Combining them, we obtain 3.286(30).
   The contribution of the uncertainty in $R$ to the 
   uncertainty in the value of $5d_{3/2}-6p_{1/2}$ matrix element is negligible. 
       \begin{table}
\caption{\label{tab6b} Contributions to the $6p_{3/2}$ tensor polarizability $\alpha_2$ in Cs 
and their uncertainties $\delta \alpha_2$ in $a^3_0$.
 The values of corresponding matrix elements $d$ (in a.u.), their sources and uncertainties $\delta d$  (in \%) also given. }
 \begin{ruledtabular}
\begin{tabular}{lrrcrr}
  \multicolumn{1}{c}{} &
 \multicolumn{1}{c}{$d$} &
\multicolumn{1}{c}{$\delta d$} &
\multicolumn{1}{c}{} &
\multicolumn{1}{c}{$\alpha_2$} &
\multicolumn{1}{c}{$\delta \alpha_2$} \\ 
\hline
  $6p_{3/2}- 5d_{3/2}$&  3.286 &         0.9&\protect~\cite{5dT}\footnotemark[1] &    114.2&  2.1\\
  $6p_{3/2}- 5d_{5/2}$&  9.916 &         0.3&\protect~\cite{5dT} &   -251.1& -1.8\\
  $6p_{3/2}- 6s      $& -6.324 &         0.1&\protect~\cite{6s6p}&    124.7&  0.2\\
  $6p_{3/2}- 6d_{3/2}$& -2.053 &         4.6&SDpT        &    11.4 &  1.0\\
  $6p_{3/2}- 6d_{5/2}$& -6.010 &         4.3&SDpT        &   -24.2 & -2.1\\
  $6p_{3/2}- 7s      $& -6.473 &         0.5&\protect~\cite{al}  &  -225.3 & -2.3\\
  $6p_{3/2}- 7d_{3/2}$& -0.969 &         1.5&SDpT       &     1.9 &  0.1\\
  $6p_{3/2}- 7d_{5/2}$& -2.868 &         1.4&SDpT        &    -4.2 & -0.1\\
  $6p_{3/2}- 8s      $& -1.462 &         0.6&\protect~\cite{al}    &    -6.2 & -0.1\\
  $6p_{3/2}- 9s      $&  0.774 &         0.6&\protect~\cite{al}    &    -1.4 &  0.0\\
$\alpha_{\text{tail}}$ &       &            & DHF       &   -7.0 &  2.1\\
Total                 &       &            &                       &   -267.3& 4.7
\end{tabular}
\end{ruledtabular}
\noindent \footnotetext[1]{Derived from the experimental $5d_{3/2}$ lifetime \protect~\cite{5dT} using  theoretical ratio
of the $6p_{3/2}- 5d_{3/2}$ and  $6p_{1/2}- 5d_{3/2}$ matrix elements. } 
 \end{table} 
       \begin{table*}
\caption{\label{tab7} Calculated and experimental values of Cs polarizabilities,
in $a^3_0$. Calculation (a) uses $5d-6p$ matrix elements data derived from the 
 $5d$ lifetime experiment \cite{5dT} (results of Tables~\ref{tab6a} and \ref{tab6b});
calculation (b) uses  $5d-6p$ theoretical all-order values (SD scaled data). 
All other contributions in calculations (a) and (b) are the same.}
 \begin{ruledtabular}
\begin{tabular}{lccrrrr}
\multicolumn{1}{c}{} &
\multicolumn{2}{c}{Present} &
\multicolumn{1}{c}{Expt.~\protect\cite{6pH}} &
\multicolumn{1}{c}{Expt.~\protect\cite{6pT}} &
\multicolumn{1}{c}{Expt.~\protect\cite{6pHH}} \\
\multicolumn{1}{c}{} &
\multicolumn{1}{c}{(a)} &
\multicolumn{1}{c}{(b) } &
\multicolumn{3}{c}{} \\
\multicolumn{1}{c}{} &
\multicolumn{1}{c}{Expt. $5d-6p$} &
\multicolumn{1}{c}{Theory $5d-6p$} &
\multicolumn{3}{c}{} \\

\hline
$\alpha_0(6p_{3/2})-\alpha_0(6s)$ &   1322(18) & 1248    & 1264(13)  & 1240.2(24)&  & \\
$\alpha_0(6p_{1/2})-\alpha_0(6s)$ &   1006(24) & 936     &    970(9) &           & 927.35(12) \\
$\alpha_2(6p_{3/2})$              &  -267(4.7) & -261.2  &  -261(8)  &-262.4(15) &  &
\end{tabular}
\end{ruledtabular}
 \end{table*} 
    
      The scalar $\alpha_0$ and tensor $\alpha_2$ polarizabilities of 
of an atomic state $v$ are calculated using formulas 
      \begin{eqnarray}
\alpha_0^v&&=\frac{2}{3(2j_v+1)}\sum_{n}
\frac{
\langle n\|D\| v \rangle^2 }
{E_{n}-E_{v}},  \\ 
\alpha_2^v&&=4\left(\frac{5j_v(2j_v-1)}{6(j_v+1)(2j_v+1)(2j_v+3)}\right)^{1/2}\nonumber \\
&&\sum_{n}
(-1)^{j_v+j_n+1}\left\{
\begin{array}{ccc}
j_v & 1 & j_n \\
 1 & j_v & 2
\end{array}
\right\} 
\frac{
\langle n\|D\| v \rangle^2 }
{E_{n}-E_{v}},
\label{alpha2}
\end{eqnarray}
where $D$ is the dipole operator and formula for $\alpha_0$ includes
 only valence part of the polarizability.
The main contributions to the polarizability,
  $\alpha_{\text{main}}$, come from
transitions between 
 $6s$, $7s$, $8s$, $9s$, $6p$, $7p$, $8p$, $9p$, $5d$, $6d$, and $7d$ 
 levels; the remainder, $\alpha_{\text{tail}}$, is calculated from summing
over all other valence-excited states of the system (which is confined in a
sphere of radius 75~$a_0$).
 The core contribution to the scalar polarizability $\alpha_{\text{core}}=15.8~a^3_0$, is taken from \cite{jkh},
 where it was calculated in random-phase-approximation  (RPA). We note that this value includes the 
 contribution from the valence shell and, therefore, must be compensated by the additional term, 
 $\alpha_{\text{vc}}$, which is equal to the contribution from the valence shell divided by $(2j_v+1)$
 with an opposite sign.
  We find that the $\alpha_{\text{vc}}$ term is negligible for $np$ states and very small (below 0.2\%)
  for the $6s$ state. 
   We list the contributions to Cs $6p$ polarizabilities in Tables~\ref{tab6a} and \ref{tab6b}.
   The corresponding electric-dipole matrix elements $d$, their sources, and uncertainties $\delta d$ are also given. 
   The values for $6s-np$ and $7s-np$ transitions are taken from Ref.~\cite{al}, where
   the ``best value'' set of these matrix elements was compiled for the calculation of the
   tensor transition polarizability $\beta$. The $6p-6d$ and $6p-7d$ matrix elements are from the present {\it ab initio}
   SDpT calculation. The values of the $5d-6p$ matrix elements are 
   derived from the 
 $5d$ lifetime experiment \cite{5dT}. The same data set is used in both tables.
 The uncertainties of all contributions are listed separately.
 The uncertainties listed in Ref.~\cite{al} are used for $6s-np$ and $7s-np$ transitions; the difference between 
 SD and SDpT data is taken to be the uncertainty of the $6p-6d$ and $6p-7d$ matrix elements calculated in this work. The
 uncertainties of the $5d-6p$ matrix elements obtained from the lifetime experiment \cite{5dT} are derived above.  
 The uncertainty of the core term $\alpha_{\text{core}}$ is taken to be 2\% based on the comparison of 
 RPA data for closed core systems with experiments and high-precision calculations. 
 The uncertainty of the remaining contribution $\alpha_{\text{tail}}$ is estimated to be 30\% 
 based on the comparison of the DHF results with correlated values. 
 
We also calculate the scalar polarizability of the $6s$ state using the same methods and data set
as for the $6p$ polarizability. The resulting value  $\alpha_0(6s)=398.2(0.9)~a^3_0$ and its uncertainty 
are dominated by contributions of the $6s-6p$ matrix element taken from experiment of Ref.~\cite{6s6p}. We use this result when calculating differences of $6p$
and $6s$ polarizabilities. The recent measurement of the ground state polarizability in Cs
yielded the value $\alpha_0(6s)=401.0(0.6)~a^3_0$ \cite{pole}. 

We compare the final results for the 
  differences of the $6p$ and $6s$ scalar polarizabilities $\alpha_0$ and the tensor polarizability 
  $\alpha_2$ with experiment in Table~\ref{tab7}. The results of the above calculation (data from Table~\ref{tab6a}, \ref{tab6b}), where we used $5d-6p$ matrix elements derived from the $5d$ lifetime experiment are listed in column (a).
   We find that the difference of the $6p_{3/2}$ and $6s$ scalar polarizabilities 
 which uses numbers for $5d-6p$ matrix elements derived from \cite{5dT} $5d$
 lifetime measurements 
 $
 \alpha_0(6p_{3/2})-\alpha_0(6s)=1322(18)~a^3_0
 $
 is inconsistent with both experimental values 1240.2(24)~$a^3_0$
 \cite{6pT} and 1264(13)~$a^3_0$ \cite{6pH}. The difference with first value, which has the smaller
 uncertainty is 4.5$\sigma$ and the difference with the second value is 2.6$\sigma$.
  The difference of the $6p_{1/2}$ and $6s$ scalar polarizabilities 
 which uses numbers for $5d-6p$ matrix elements derived from \cite{5dT} $5d$
 lifetime measurements 
  $
 \alpha_0(6p_{1/2})-\alpha_0(6s)=1006(24)~a^3_0
 $
 is also inconsistent with the most recent and most precise experimental value 927.35(12)$~a^3_0$
 \cite{6pHH} by 3.2$\sigma$. The value for the $6p_{3/2}$ 
 tensor polarizability  $-267.3(4.7)$~$a^3_0$ has much larger uncertainty owing to strong cancellation of the contributions from 
 different transitions, and the difference is 1$\sigma$. We note that if we were to use another $5d_{5/2}$ lifetime experiment
 \cite{5dH}, the discrepancies with polarizability measurements only increase. Thus, neither $5d_{5/2}$ lifetime experiment
 \cite{5dT,5dH} is consistent  with either \cite{6pH}, \cite{6pT}, or \cite{6pHH} Stark shift measurements within the quoted uncertainties.  

We calculate that the experimental value of   $\alpha_0(6p_{1/2})-\alpha_0(6s)=927.35(12)$~$a^3_0$
\cite{6pHH} corresponds to the lifetime of the $5d_{3/2}$ state $\tau_{5d_{3/2}}=975(14)$~ns  and 
the experimental value of   $\alpha_0(6p_{3/2})-\alpha_0(6s)=1240.2(2.4)$~$a^3_0$
\cite{6pT} corresponds to the lifetime of the $5d_{5/2}$ state $\tau_{5d_{5/2}}=1359(18)$~ns.
The uncertainties in these lifetime values are dominated by the uncertainties in the values of $6p-6d$
transitions and the uncertainty in the contribution $\alpha_{\text{tail}}$ as evident from 
Table~\ref{tab6a}. 

 Finally, we repeated the polarizability calculation by replacing the $5d-6p$ matrix elements derived from the 
 lifetime experiment by our theoretical values (SD$_{\text{sc}}$) from Table~\ref{tab1}. All other matrix elements 
 and contributions are exactly the same as in the first calculation. The results are listed in 
 column (b) of Table~\ref{tab7}. As expected, they are quite different from the previous calculation (a)
 as our theoretical $5d-6p$ matrix elements are substantially different from the values derived from 
 $5d$ lifetimes. We find that our theoretical polarizability data are in good 
 agreement (0.4\%-1\%) with experimental results.  
 
 In conclusion, we find  the experimental measurements of $5d$ lifetime and $6p$ scalar polarizabilities to be inconsistent 
 within the uncertainties quoted by the experimental groups. Our theoretical calculations are consistent with 
 polarizability experiments but not with the lifetime measurements. 
Thus, further measurements of the properties of $5d$ and $6p$ states are of great interest for clarification of this
issue and for providing benchmark values for $5d-6p$ matrix elements. 
 
\end{document}